\documentclass[aip, prb, reprint, groupedaddress, floatfix]{revtex4-1}

\usepackage{graphicx}  
\usepackage{dcolumn}   
\usepackage{bm}        
\usepackage{amssymb}   
\usepackage{amsmath}
\usepackage{upgreek}

\begin{document}

\title{Surface participation and dielectric loss in superconducting qubits}

\author{C. Wang}
\author{C. Axline}
\author{Y. Y. Gao}
\author{T. Brecht}
\author{Y. Chu}
\author{L. Frunzio}
\author{M. H. Devoret}
\author{R. J. Schoelkopf}
\affiliation{Department of Applied Physics and Physics, Yale University, New Haven, CT 06520, USA}

\date{\today}
 
\begin{abstract}
We study the energy relaxation times ($T_1$) of superconducting transmon qubits in 3D cavities as a function of dielectric participation ratios of material surfaces.  This surface participation ratio, representing the fraction of electric field energy stored in a dissipative surface layer, is computed by a two-step finite-element simulation and experimentally varied by qubit geometry.  With a clean electromagnetic environment and suppressed non-equilibrium quasiparticle density, we find an approximately proportional relation between the transmon relaxation rates and surface participation ratios.  These results suggest dielectric dissipation arising from material interfaces is the major limiting factor for the $T_1$ of transmons in 3D cQED architecture.  Our analysis also supports the notion of spatial discreteness of surface dielectric dissipation.
\end{abstract}


\maketitle


Circuit quantum electrodynamics (cQED) systems have emerged as promising platforms for quantum information processing, powered by dramatic improvement of the coherence times of superconducting qubits over the past decade~\cite{Devoret2013}.  Such an improvement has been the result of collective efforts in multiple aspects~\cite{Martinis2014}, such as suppression of charge noise and flux noise~\cite{Koch2007}, better control of the electromagnetic environment~\cite{Paik2011}, elimination of deposited dielectric materials~\cite{Martinis2014}, development in surface treatment~\cite{Barends2013}, dilution of surface effects by expanding field volume~\cite{Paik2011}, and improved filtering and shielding against stray radiation\cite{Corcoles2011}.  However, it has been difficult to quantify how much each of these individual measures contribute to the overall improvement.  As a result, it remains elusive what the dominant limiting factors are for the coherence of state-of-the-art superconducting qubits such as the 3D and planar transmons.  

The superior lifetimes ($T_1$) of qubits with larger footprints \cite{Paik2011} or with more advanced surface preparation~\cite{Barends2013} strongly suggest the important role of dielectric dissipation~\cite{Martinis2005} from material surfaces.  In this letter, we quantitatively extract surface dielectric dissipation in transmon qubits through a combined experimental and numerical study.  We find that surface dielectric dissipation is probably still the major limiting factor for $T_1$ of transmons in 3D cQED architecture, and so far there is no indication of additional loss mechanisms (up to the level of $Q\sim10^7$) under our experimental condition.  Our analysis also indicates that surface loss for a sub-micrometer area cannot be captured by a uniform loss tangent model, consistent with the hypothesis of discrete dissipation from a small number of microscopic two-level states (TLS)~\cite{Martinis2005, Gao2008, Holder2013, Grabovskij2012, Sarabi2015}.


Relaxation of superconducting qubits or resonators can be caused by many dissipative channels such as dielectric loss, conductive loss, and radiation into free space~\cite{Martinis2014}.  Dielectric loss can be further decomposed into contributions from various materials or components, so that:
\begin{equation}
\frac{1}{T_1}=\frac{\omega}{Q}=\omega \sum_i \frac{p_i}{Q_i} + \Gamma_0
\end{equation}
where $T_1$, $Q$ and $\omega$ are the relaxation time, quality factor (for energy decay) and angular frequency of the qubit or resonator, $\Gamma_0$ is the relaxation rate induced by non-dielectric channels, $Q_i = 1/\tan\delta_i$ is the quality factor of the $i^{th}$ material with a dielectric constant of $\epsilon_i$ (with $\tan\delta$ known as the loss tangent), and $p_i$ is its participation ratio defined as the fraction of electric field energy stored within the volume of this material. 

Crystalline substrates of cQED devices often store a large fraction of electric field energy ($p_i\sim90\%$), but reportedly show very small loss tangent ($\tan\delta_i<10^{-6}$ for bulk sapphire~\cite{Creedon2011} and silicon~\cite{Martinis2014}).  On the other hand, if a microscopic layer of contaminants such as oxide, adsorbed water or organics forms at the metal-substrate (MA), substrate-air (SA) and metal-air (MA) interfaces~\cite{Sandberg2012, Wenner2011}, they have much smaller $p_i$ but may still induce significant dissipation with a large $\tan\delta_i$  on the order of $10^{-3}$-$10^{-2}$.  Previous studies~\cite{Wang2009, Sage2011, Geerlings2012, Megrant2012} have found a positive correlation between the quality factors of planar resonators and their feature sizes which can be used to vary $p_i$.  However, a quantitative test of Eq.~(1) has been challenging due to the presence of other energy relaxation channels ($\Gamma_0$) that have not been fully under control.  

\begin{figure}[tbp]
    \centering
    \includegraphics[width=0.47\textwidth]{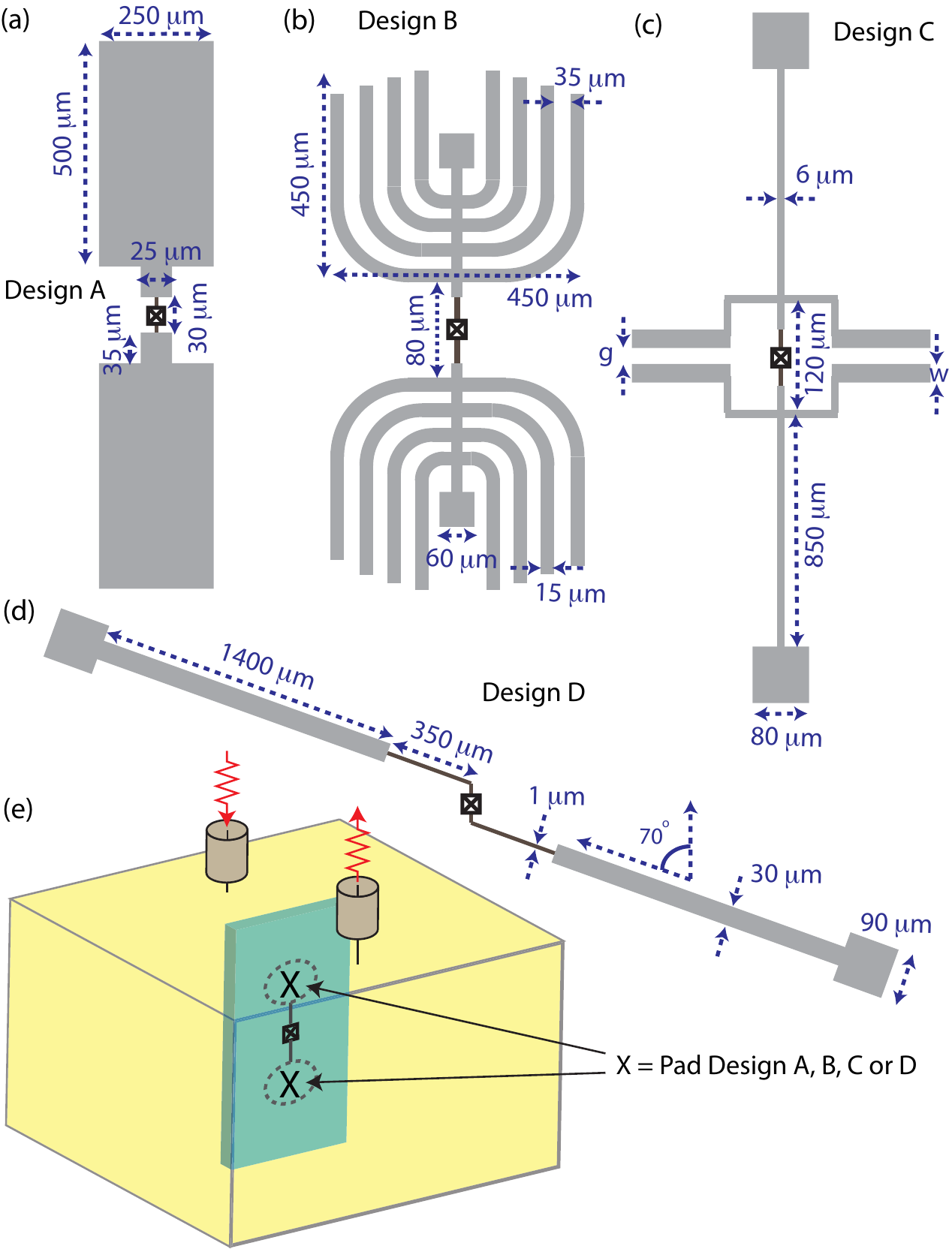}
    \caption{\textbf{Geometry of four different designs of transmon qubits used in this study.}  Most dimensions of the electrodes for each design (A, B, C, D) are fixed and labeled in respective panels (a, b, c, d).  An exception is that Design C has multiple variations with the dimensions $g=w$ ranging from 1.5 $\mu$m to 30 $\mu$m.  For convenience, we define ``leads" as the portion of the electrodes with widths 1 $\mu$m or smaller, which only appears in direct connection to the junction in all our designs, as shown in dark brown.  The rest of the electrodes wider than 1 $\mu$m are called ``pads" and shown in lighter grey.  (e) Schematic of the standard 3D cQED setup.  Transmon qubits are installed in rectangular waveguide cavities and coupled to the TE101 mode for control and readout.}
    \label{geometry}
\end{figure}

Here we study the energy relaxation time, $T_1$, of transmon qubits as a function of surface dielectric participation ratio, $p_i$.  Strong suppression of radiation loss is achieved by implementing the 3D cQED architecture~\cite{Paik2011} where the 3D cavity enclosure provides a clean electromagnetic environment free of spurious modes.  The cavity Q and qubit-cavity detuning are sufficiently large to avoid any appreciable Purcell effect.  Qubit relaxation due to non-equilibrium quasiparticles can be estimated and suppressed by monitoring and controlling quasiparticle decay time~\cite{Wang2014, Vool2014, QPnote}.  Furthermore, transmons are less sensitive to vortex ac loss than linear resonators because most inductive energy is stored in the Josephson junction rather than the electrodes subjected to vortex penetration.  Suppression of these relaxation channels allows us to vary the qubit geometry to change $p_i$ by more than an order of magnitude, making quantitative comparison of surface dielectric loss in different devices viable. 

Each qubit in this study is composed of a single Al/AlO$_x$/Al Josephson junction and a pair of electrodes forming a shunting capacitor.  We report $T_1$ measured with standard techniques for four different geometric designs of transmons as shown in Fig.~1.  All devices are fabricated on sapphire substrates with identical processes of shadow-mask evaporation and lift-off~\cite{Supplementary}, and therefore are assumed to have the same loss tangent for the same type of surfaces.  All devices have qubit frequency $\omega/2\pi\approx6$ GHz and cavity frequency $\omega_c/2\pi\approx9$ GHz.


\begin{figure}[tbp]
    \centering
    \includegraphics[width=0.48\textwidth]{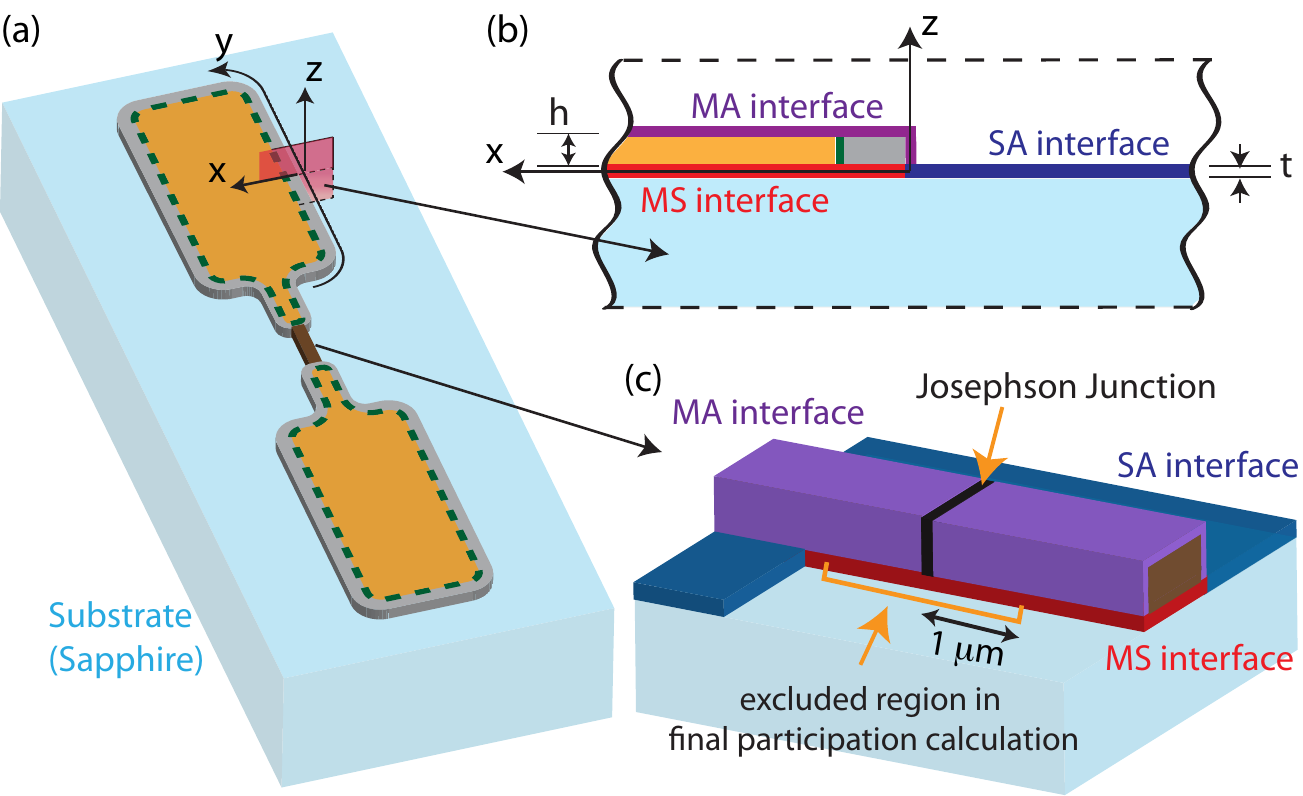}
    \caption{\textbf{Illustration of the two-step simulation strategy for computing surface participation ratios.}  (a) Schematic of a transmon qubit with its electrodes color-coded into several regions.  Grey and yellow represents the perimeter and the interior regions of the wide pads of the electrodes respectively, and brown represents the narrow leads.  A global coarse 3D simulation can accurately determine the electric fields across the yellow region, but not in the grey and brown region near the edge of the metal.  (b) A cross-section view of the electrode near a metal edge.  The electric field distribution within this plane can be computed by a fine 2D simulation.  (c) A simplified schematic of the region near the Josephson junction, which is simulated by a local fine 3D simulation.  The MA, SA and MS interfaces are defined in (b) and (c).  In our final account of surface participation ratios, contribution from the region within 1 $\mu$m from the junction is excluded.  All drawings are not to scale.}
    \label{3Dsim}
\end{figure}

Full electromagnetic simulation of surface participation ratio of transmon qubits faces significant numerical challenges due to the large span of length scales.  One may attempt to model transmon electrodes and any dissipative interface layers as 2D films, and infer $p_i$ from a surface integral of electric field energy.  However, such an integral is divergent towards the edge of the films \cite{Jackson}.  This divergence is avoided only when the material thicknesses are fully accounted for, as was done in a cross-sectional simulation of transmission line resonators~\cite{Sandberg2012, Wenner2011}.  Without a similar translational symmetry, a proper calculation of $p_i$ for a transmon qubit generally requires simulation of 3D field distribution in mm-sized space with sub-nm resolution in critical regions, far exceeding practical computation capacities. 

\begin{figure*}[tbp]
    \centering
    \includegraphics[width=0.76\textwidth]{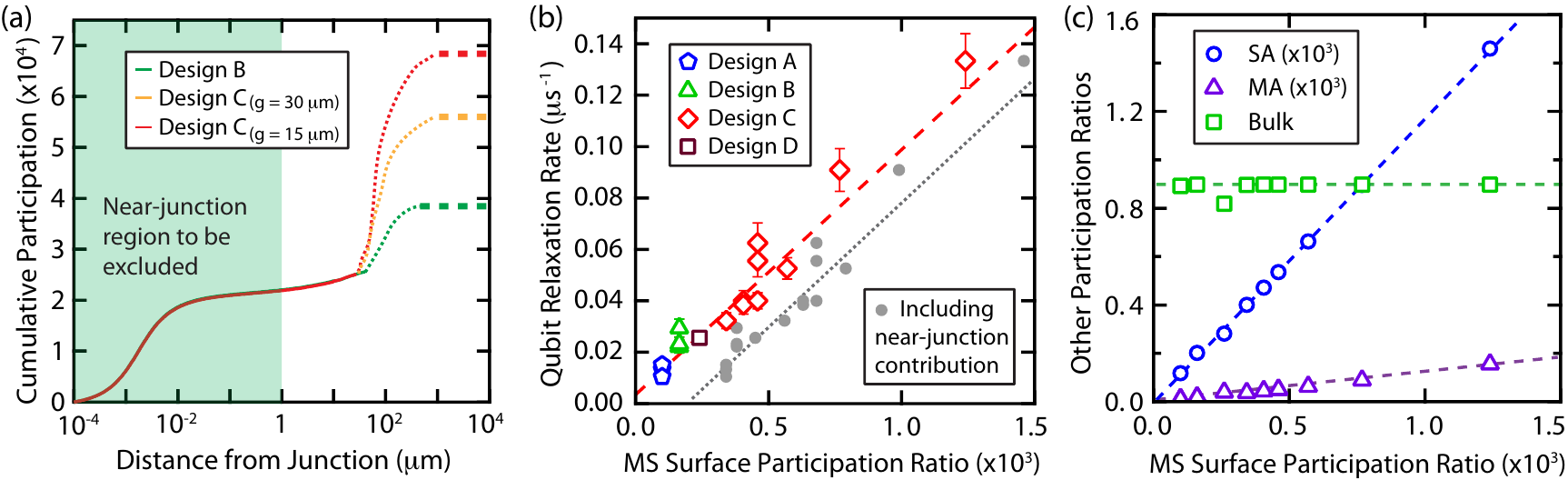}
    \caption{\textbf{Surface participation and qubit lifetime.}  (a) Spatial distribution of simulated surface participation contribution, shown for MS interface of selected transmon designs.  Solid curves show cumulative integral of the MS surface participation ($p_{MS}$) from the electrode leads as they extend from the junction in distance (horizontal axis), indicating total surface participation from the junction to that point. The thick dashed lines indicate the total $p_{MS}$ of all features.  The dotted lines notionally represent contribution from the electrode pads as a whole. 
  (b) Open symbols show measured transmon $1/T_1$ as a function of $p_{MS}$ excluding contribution from the sub-micron ``near-junction region" (green shaded area in (a)) that most probably contains no TLS.  Red dashed line is a fit to Eq.~(1).   The same set of data including the near-junction contribution is plotted as grey filled circles, with corresponding fit to Eq.~(1) shown as the dotted line.   (c) SA, MA surfaces and substrate bulk participation ratios ($p_{SA}$, $p_{MA}$ and $p_{bulk}$) as a function of $p_{MS}$ for transmon devices in this study.  Dashed lines are guides to the eye.}
    \label{plot}
\end{figure*}

To overcome the numerical challenges, we employ a two-step simulation technique by combining a coarse 3D simulation of the entire qubit-cavity system [Fig.~2(a)] and fine simulations of representative local regions [Fig.~2(b, c)].  A significant part of the surface participation is associated with regions with highly concentrated electric field such as the edges of the electrodes and the leads near the junction.  We argue that the electric field distribution in these regions should have a local scaling property independent of the electromagnetic boundary conditions far away.  These scaling properties can be obtained from simulations of local regions with sub-nm resolution and subsequently applied to the global simulation to compute the surface participation ratios~\cite{Supplementary}.  We assume thicknesses of $t=3$ nm and dielectric constants of $\epsilon=10$ for all lossy interfaces for easy comparison with a previous simulation of planar resonators~\cite{Wenner2011}.  Using different assumptions here would rescale the participation ratios but not change our conclusions qualitatively.  


Our simulation shows that a significant contribution to surface participation arises from the region around the junction leads less than 100 nm away from the junction itself [Fig.~3(a)].  This contribution is mostly independent of electrode geometry, and can be dominant for devices with relatively small surface participation~\cite{Supplementary}.  However, if surface dielectric dissipation originates from a discrete set of TLS with density similar to junction defects~\cite{Martinis2005, Shalibo2010, Schreier2008, Stoutimore2012} ($\sim1$ $\mu$m$^{-2}$GHz$^{-1}$), it is most likely that such a small volume of macroscopically lossy material contains no resonant TLS and thus appear dissipationless.  This motivates us to introduce a dimensional cutoff and exclude the participation contribution from this near-junction region.  We choose to set this cutoff at a distance of 1 $\mu$m from the junction, but any choice on the order of $100$ nm to $10$ $\mu$m does not affect the total participation significantly because the participation contribution from this intermediate region of the electrode leads is insignificant [Fig.~3(a)]. The resultant total $p_{MS}$ from the rest of the MS surface is approximately proportional to the measured $1/T_1$ for all our devices [Fig.~3(b)].  Similarly, we also observe $p_{MA}$ and $p_{SA}$ proportional to $1/T_1$.~\cite{Supplementary}

The proportionality between qubit decay rate and surface participation ratios strongly suggests surface dielectric loss as the dominant relaxation mechanism for all transmons in this study.  Based on Eq.~(1), any geometry-independent dissipation mechanism is expected to induce a constant relaxation rate $\Gamma_0$ to all our devices.  If we were to include the near-junction contribution (as noted above) in $p_{MS}$, a linear fit of our data to Eq.~(1) would produce an unphysical negative y-interception [Fig.~3(b)].  This reinforces the notion of spatial discreteness of surface loss and the necessity of a cutoff.   After implementing the cutoff, we see a very small residual qubit decay rate ($3\pm1$ ms$^{-1}$), which can be fully explained by the magnitude of quasiparticle dissipation and vortex ac loss as we noted previously.  Therefore there is no evidence of any geometry-independent loss mechanisms, such as from the crystalline substrate or the Josephson junction itself, that limit transmon lifetimes on the level of $Q\sim10^7$.  The absence of loss from the junction may be a result of the small junction size (0.04 $\mu$m$^2$) so that no resonant junction defects are encountered in this study.  We also note that surface loss mechanisms consistent with our observed geometric scaling should not be viewed strictly due to impurity or defect-like TLS.  Potential alternative mechanisms closely related to surface electric field energy, such as phonon radiation due to surface piezo-electricity~\cite{Ioffe2004, Gustafsson2014}, may also be broadly included in the surface dielectric loss in this analysis.

We cannot determine which of the three surfaces are the dominant contributor based on these data alone, because all three participation ratios change approximately in proportion when the qubit geometry is varied [Fig.~3(c)].
We can determine a weighted sum of the loss tangents of the three surfaces, $\tan\delta_{MS}+1.2\tan\delta_{SA}+0.1\tan\delta_{MA}=(2.6\pm0.1)\times10^{-3}$.   To extend our analysis to distinguish different interfaces, one generally needs to go beyond a planar layout of transmon electrodes, for example by incorporating striplines or microstrips. 

\begin{figure}[tbp]
    \centering
    \includegraphics[width=0.48\textwidth]{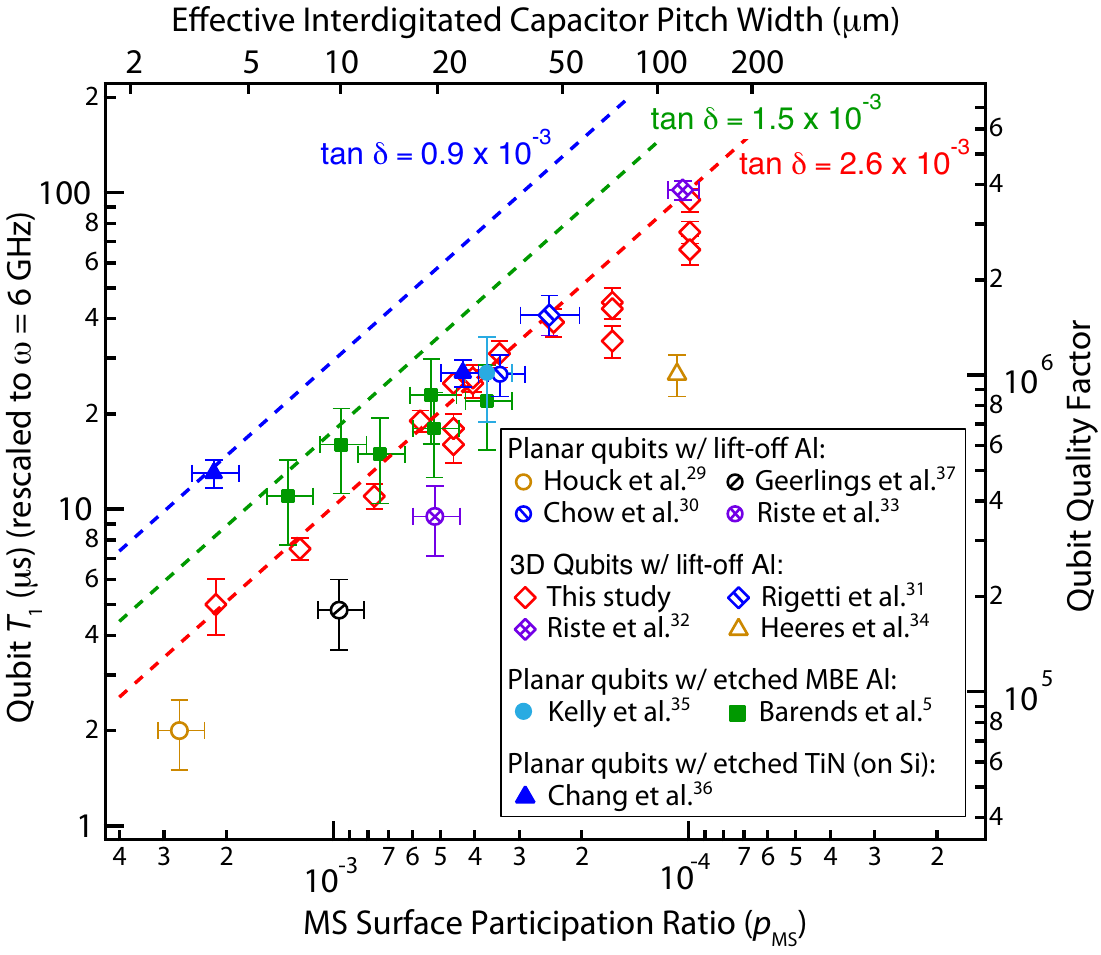}
    \caption{\textbf{Transmon lifetime vs.~MS surface participation ratio for selected literature data.}  Open symbols represent various 3D and planar transmons fabricated with the single-step aluminum lift-off process similar to this study.  Filled symbols represent transmons with electrode pads fabricated with etch processes to preserve clean MS surface.  The vertical axes represent qubit quality factors or the equivalent $T_1$ at $\omega/2\pi=6$ GHz.  The bottom axis shows $p_{MS}$ in a reversed log scale.  The top axis shows the equivalent interdigitated capacitor pitch width of a planar qubit for corresponding $p_{MS}$, a helpful alternative unit of surface participation.~\cite{Supplementary}}
    \label{mplot}
\end{figure}


We have further calculated or estimated $p_{MS}$ for reported planar and 3D transmons from the literature~\cite{Houck2008, Chow2014, Rigetti2012, Riste2013, Riste2015, Heeres2015, Kelly2015, Barends2013, Chang2013, GeerlingsThesis}, and Fig.~4 shows the Q factors or $T_1$'s of some of these devices as a function of $p_{MS}$.  All data points with a single-step aluminum lift-off process similar to ours fall near or below the surface-loss line of $\tan\delta=2.6\times10^{-3}$ (red dashed line), consistent with the surface dielectric loss determined in this study.  We believe similarly-fabricated qubits performing substantially worse than this surface-loss line are limited by other mechanisms.  Early generation of planar transmons may incur losses due to non-equilibrium quasiparticles or lossy components of the device package~\cite{Houck2008}, and the 3D ``vertical" transmons~\cite{Heeres2015} may be severely limited by conduction loss across the cavity seam~\cite{Brecht2015}.  

Several recent studies used subtractively-patterned MBE aluminum~\cite{Barends2013} or TiN~\cite{Chang2013} films for transmon electrodes.  These processes were intended for preserving pristine MS interface, and subsequent improvement of $T_1$ suggests MS interface may indeed play an important role in the total surface loss.  We find several data points for these qubits (the leftmost filled symbols) above our $\tan\delta=2.6\times10^{-3}$ line at relatively high $p_{MS}$, confirming higher surface quality than have been measured in this present study.   However, these surface improvements have not been fully translated into the best possible performance for devices with lower $p_{MS}$, as indicated by their surface-loss bounds (blue and green dashed lines in Fig.~4).  It suggests the presence of other dissipation channels yet to be fully suppressed in these high-material-quality planar qubits.  These devices also include shadow-mask evaporated junction leads with lower quality surfaces that can have appreciable surface participation and limit qubit $T_1$. 

Looking forward, further advance of coherence times of superconducting qubits will hinge on a combination of improving material surface quality and further reducing surface participation ratios.  The state-of-the-art planar transmons have implemented large-sized planar capacitors~\cite{Kelly2015, Chow2014} to reduce surface participation, yielding substantial gains in qubit lifetimes. 
One may naively expect that millimeter-sized 3D transmons may have smaller $p_i$ by orders of magnitude and make dielectric loss irrelevant.  The present study shows this is not the case.  Furthermore, our simulations find that merely engineering larger and more-separated electrodes will incur significant $p_i$ from the metal leads required to wire up the Josephson junction.  Nevertheless, substantial further reduction of surface participation in qubits can be achieved by more complex three-dimensional designs such as deep-etched\cite{Bruno2015} or suspended structures~\cite{Brown1999}.  With no hard limit in sight, innovative low-participation designs and improved surface quality, together with modest progress in suppressing non-equilibrium quasiparticles, are expected to bring another order of magnitude increase in the lifetime of transmon qubits.

We thank R.~W.~Heeres and P.~ Reinhold for experimental assistance, Z.~Minev for helpful discussions, and the support of M. Guy of the Yale Science Research Software Core.  C.~A.~acknowledges support from the NSF Graduate Research Fellowship under Grant No.~DGE-1122492.  This research was supported by IARPA under Grant No.~W911NF-09-1-0369 and ARO under Grant No.~W911NF-09-1-0514.  Facilities use was supported by YINQE and NSF MRSEC DMR 1119826.   

\bibliography{SurfaceLoss_Final2}

\begin{thebibliography}{40}
\bibitem [{QPn()}]{QPnote}%
  \BibitemOpen
  \href@noop {} {\bibinfo  {journal} {By measuring qubit $T_1$ under conditions
  with short quasiparticle decay time, $\tau_{ss} < 1$ ms, achievable with a
  small cooling magnetic field, we limit quasiparticle-induced qubit relaxation
  to below 1/(250 $\mu$s). This assumes stray quasiparticle generation rate
  $g\approx1\times10^{-4}$/s for all our devices with the same filtering and
  shielding, as measured in Ref.~19}\ }\BibitemShut {NoStop}%
\end{thebibliography}

\end{document}


\preprint{APS/123-QED}

\title{Supplementary Materials for ``Surface participation and dielectric loss in superconducting qubits"}

\author{C. Wang}
\author{C. Axline}
\author{Y. Y. Gao}
\author{T. Brecht}
\author{Y. Chu}
\author{L. Frunzio}
\author{M. H. Devoret}
\author{R. J. Schoelkopf}
\affiliation{Department of Applied Physics, Yale University, New Haven, Connecticut 06511, USA}

\date{\today}


\maketitle


\section{Simulation Methods for Surface participation}

Participation ratios embody a convenient method to account for dissipative loss in dielectric systems.  The participation ratio of a certain material or component of the circuit can be calculated by integrating electric field energy in an electromagnetic simulation of the exact model of the device.  However, from typical adaptive-mesh simulation techniques, it is very difficult to produce convergent values of total field energy stored in thin surface layers in a 3D qubit-cavity system due to the disparity of length scales.  To address this challenge, we introduce a two-step simulation technique to calculate participation ratios for three different material surfaces---metal-substrate (MS), substrate-air (SA), and metal-air (MA)---for a variety of 3D transmon qubit designs.  The results of simulated surface participation ratios and measured lifetimes for these qubits (used for Fig.~3 and 4 of the main text) are listed in Table S1.  In this section, we describe the simulation methods used to obtain these results.

\begin{table*}[tbp]
\caption{\textbf{Results of simulated surface participation ratios and measured lifetimes of transmon qubits.}  Participation ratios are multiplied by $10^{-4}$, and calculated by summing over contributions from various regions of the surfaces as shown for $p_{\textrm{MS}}$ for example.  Starred ``$p_{i}$ total" excludes contribution from the region within 1 $\upmu$m from the junction.  Measured $T_1$'s are listed for individual devices, with the uncertainty representing one standard deviation of its fluctuation over time.
}

\centering  
\begin{tabular}{c c c c c c c c c c c} 
\hline\hline                        
		& $p_{\textrm{MS,near}}$	& $p_{\textrm{MS,mid}}$&$p_{\textrm{MS,far}}$&$p_{\textrm{MS,per}}$&$p_{\textrm{MS,int}}$&$p_{\textrm{MS}}$	& $p_{\textrm{MS}}$	& $p_{\textrm{SA}}$	&$p_{\textrm{MA}}$& 		\\
Design	& leads 		& leads 		& leads 		& pads 		& pads		& \:\:total\:\:		& \:\:total*\:\:
		& \:\:total*\:\:		& \:\:total*\:\:	& $T_1$($\upmu$s)	\\
		& ($<1\upmu$m)	&($1-10\upmu$m)&($>10\upmu$m)&perimeter	&interior	\\
\hline
A		& 2.4		& 0.16	& 0.01	& 0.51	& 0.32	& 3.4	& 0.99	& 1.18	& 0.11	& 75$\pm$6, 66$\pm$7, 95$\pm$8 \\
B		& 2.2		& 0.20	& 0.19	& 0.88	& 0.37	& 3.8	& 1.64	& 2.02	& 0.19	& 34$\pm$4, 45$\pm$5, 43$\pm$3 \\
C$_{30}$	& 2.2	& 0.17	& 0.09	& 2.33	& 0.82	& 5.6	& 3.41	& 4.01	& 0.37	& 31$\pm$3\\
C$_{20}$  	& 2.2	& 0.18	& 0.09	& 2.83	& 0.95	& 6.3	& 4.05	& 4.72	& 0.44	& 25$\pm$2.5, 26$\pm$2.5\\
C$_{15}$  	& 2.2	& 0.17	& 0.09	& 3.37	& 1.05	& 6.8	& 4.59	& 5.36	& 0.50	& 25$\pm$2, 16$\pm$2, 18$\pm$2\\
C$_{10}$  	& 2.2	& 0.17	& 0.09	& 4.23	& 1.19	& 7.9	& 5.69	& 6.63	& 0.64	& 19$\pm$1.5\\
C$_{6}$  	& 2.2	& 0.17	& 0.09	& 6.05	& 1.35	& 9.9	& 7.67	& 8.96	& 0.89	& 11$\pm$1\\
C$_3$		& 2.2	& 0.17	& 0.09	& 11.1	& 1.17	& 14.6	& 12.4	& 14.6	& 1.6	& 7.5$\pm$0.6\\ 
C$_{1.5}$	& 2.2	& 0.17	& 0.09	& 19.7	& 1.42	& 23.6	& 21.4	& 23.4	& 3.2	& 5$\pm$1\\ 
D		& 2.1		& 0.19	& 1.27	& 0.58	& 0.36	& 4.5	& 2.40	& 2.82	& 0.41	& 39$\pm$4\\
\hline 
\end{tabular}
\label{table}
\end{table*}

We use a commercial high-frequency electromagnetic solver (Ansys HFSS) to simulate the entire qubit-cavity system on a $\upmu$m-to-mm scale [Fig.~S1(a)], where the aluminum film and surface dielectric layers are modeled as 2D sheets with zero thickness.  The Josephson junction and the aluminum leads very close to the junction (within 1 $\upmu$m) are modeled as a lumped element. This simulation is carried out at the qubit frequency, and similar to those routinely done for black-box quantization of cQED systems~\cite{Nigg2012}.  It provides the overall electric field distribution on a coarse scale ($\sim\upmu$m), but does not accurately reflect the highly-concentrated fields at electrode edges or near narrow leads approaching the junction that are critical to the total surface participation ratios.  To take into account the field distribution in these regions and supplement the global simulation, we perform additional local electrostatic simulations (using Ansys Maxwell) with sub-nm resolution. 


For convenience, we divide the surface dielectric layers in a transmon qubit into two regions: 1) those associated with the large ``pads,'' metal traces $>1$ $\upmu$m wide intended to form the external shunting capacitor of the transmon, and 2) those associated with the narrow ``leads,'' metal traces $\leq1$ $\upmu$m wide that are used to wire-up the junction with the pads.  Such definitions are straightforward for MS and MA surfaces in direct contact with the electrodes.  For the SA surface, we associate SA dielectric within 1 $\upmu$m of lead edges with the ``leads,'' and the remainder with the ``pads.''

In this study we mostly vary the geometry of the pads to vary the total surface participation ratio. The observed changes in $T_1$, largely correlated with pad surface participation, highlight their importance.  However, in our analysis we also explicitly calculate surface participation from the leads. This contribution has not been considered before in various implicit applications of surface participation analysis of planar resonators to Josephson-junction qubits~\cite{Barends2013, Chang2013}.



\subsection{Surfaces associated with the electrode pads}

The electrode pads are the large structures of the transmon qubits that determine qubit-cavity coupling and are often close to a millimeter in size.  Despite their large area, the majority of the electric field energy stored in their associated surfaces exists near the edges of the pads. Approximation of the electrode pads and surface dielectric layers as 2D sheets, a necessary step in full-scale simulations, results in divergent integrals for total field energy at these edges.

\begin{figure}[tbp]
  \centering
  \includegraphics[width=0.65\linewidth]{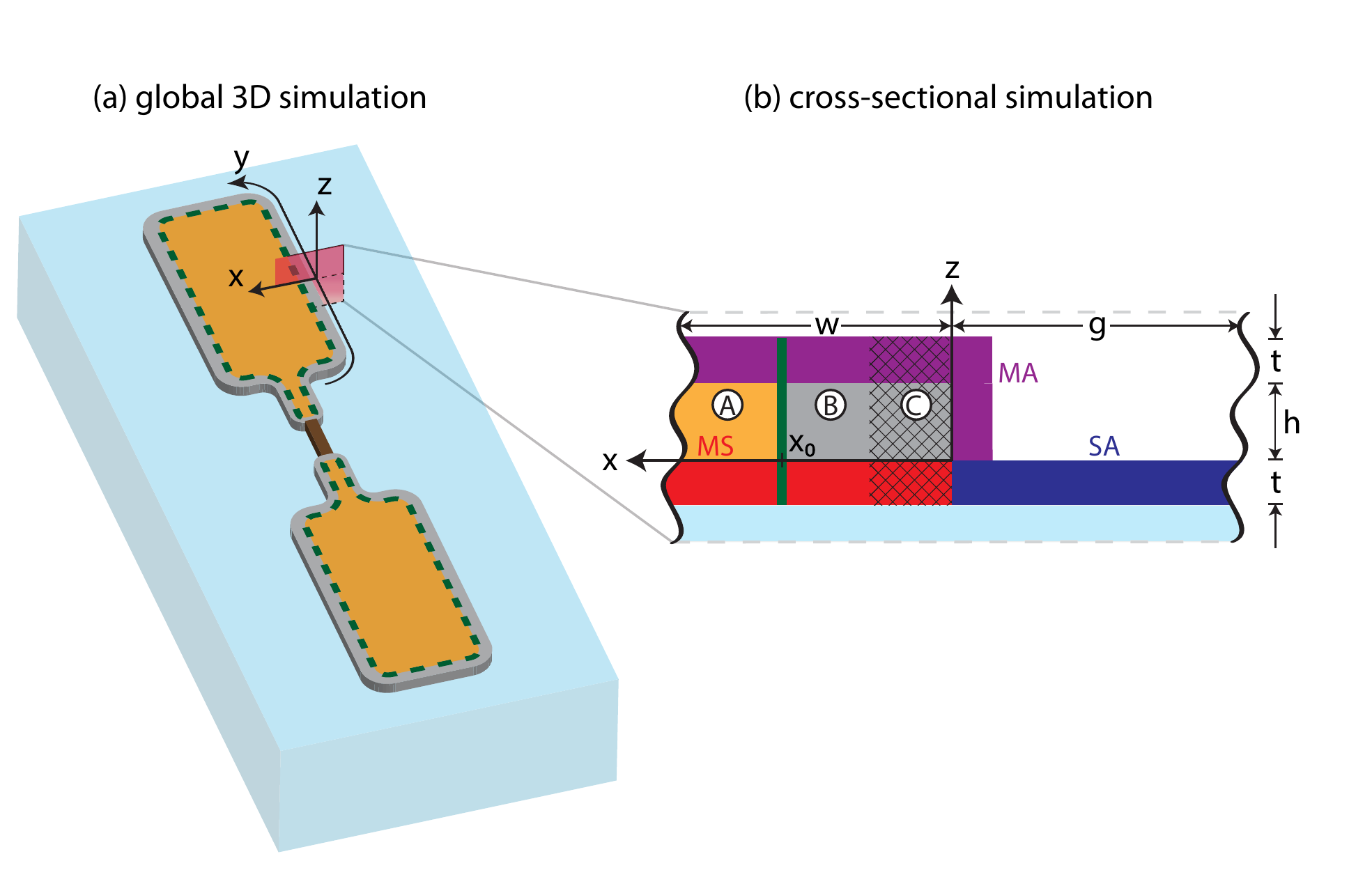}
  \caption{(a) The global high-frequency 3D simulation includes features in the entire centimeter-scale cavity, such as substrate, pads, and most of the leads. Conductors are simulated as perfectly-conducting sheets, and interface layers are omitted. The inner (orange) and perimeter (gray) regions, described in Section A, are separated by the dashed green contour. The red window shows the placement of the cross-section in (b). (b) This cross-section of the edge of an electrode pad on the substrate (light blue) labels the geometry within the 2D electrostatic simulation that supplements the global simulation along pad edges. The three interfaces of interest with thickness $t$ are shown in red (MS), purple (MA), and blue (SA), while the two regions of the superconductor (thickness $h$) are shown in orange (\textcircled{A}, ``interior region'') and gray (``perimeter region''). The perimeter region is divided into a cross-hatched region \textcircled{C}, which fails to converge in the global simulation, and a region \textcircled{B} that is convergent in both simulations. The division between \textcircled{A} and \textcircled{B} occurs at $x=x_0$. When simulated, symmetric boundaries are established to represent an interdigitated capacitor (IDC) style device with conductor width $w$ and gap width $g$ ($w,g\ll x_0$). All dimensions are not to scale.}
     \label{areas}
\end{figure}

To avoid this divergence, we first divide the electrode pads and the associated MS and MA surfaces into ``perimeter regions'' and ``interior regions'' [Fig.~\ref{areas}(a,b)] with their boundary set at a constant distance ($x_0$, typically 1 $\upmu$m) from the edge.  (The SA surface can be similarly divided by a contour at a constant distance $x_0$ from the outside of the edge.  The treatment of the SA surface is otherwise analogous to that of MS.)  In a global coarse 3D simulation, electric field in the interior regions does not have sharp variations, and therefore easily converges to spatial distributions that we may immediately record as $\mathbf{E}_{\textrm{MA}}(x,y)$ and $\mathbf{E}_{\textrm{MS}}(x,y)$ at the top and bottom surfaces of the electrode pads respectively.  We use these field distributions to calculate the surface participation associated with the interior region of the pads (denoted by the subscript ``int''):
\begin{equation}
p_{i,\textrm{int}} = t\iint_{\textrm{int}}{\frac{\epsilon}{2} |\mathbf{E}_i(x,y)|^2 dxdy}/U_{\textrm{tot}}
\end{equation}
where $i$ = MS or MA, and $U_{\textrm{tot}}$ is the total electric field energy in the entire space (dominated by energy in the substrate and vacuum).  Here we have multiplied the field integral by the assumed thickness of the surface layer, $t$ = 3 nm, further assuming that the electric field is uniform across that thickness.

The perimeter regions can be described by a spatial coordinate $(x, y, z)$ as shown in Fig.~\ref{areas}(a,b), where the $y$-axis winds around the edges of the pads, remaining tangent. We further divide the perimeter regions into two halves.  Energy in the half adjacent to the edge $(0<x<x_0/2)$ fails to converge, regardless of initial mesh parameters, following the adaptive mesh refinement process. The other half $(x_0/2<x<x_0)$ can be made to converge using mesh parameters that are computationally accessible. The key concept to our strategy is to employ a constant ratio, or ``scaling factor" $F_i$, to convert the integrated field energy in the convergent half into that of the entire perimeter regions, so that
\begin{equation}
p_{i,\textrm{per}} = F_i t\int_{x_0/2}^{x_0}dx\oint_{y}{\frac{\epsilon}{2}|\mathbf{E}_i(x,y)|^2 dy}/U_{\textrm{tot}}
\label{piper}
\end{equation}

\begin{figure}[tbp]
  \centering
  \includegraphics[width=0.76\linewidth]{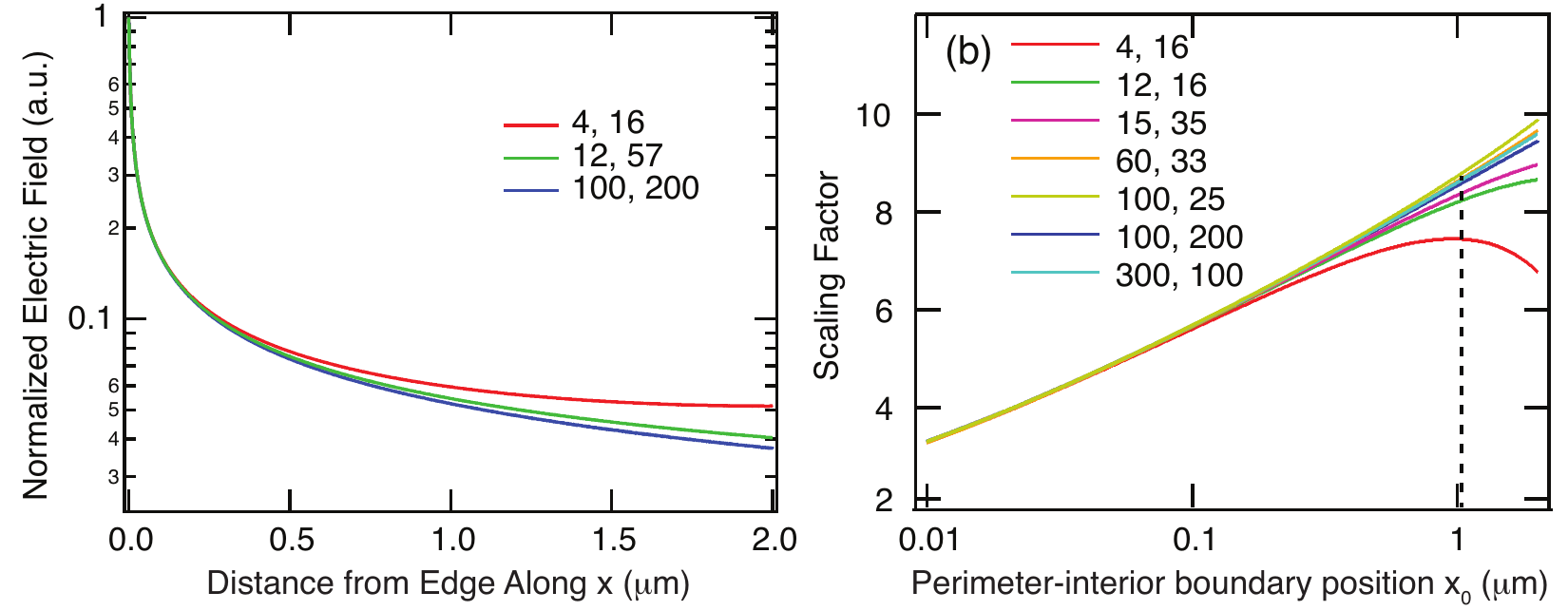}
  \caption{(a) Normalized distribution of electric field within a cross section of the MS interface near a metal edge, $f(x,-t/2)$.  It is calculated from the edge inward along a line that bisects the MS interface [Fig.~\ref{areas}(b)]. Values are plotted for simulations of three sets of boundary conditions, where the first number represents the width of the conducting feature $w$, and the second number is the distance between features, $g$, as in Fig.~\ref{areas}(b). (b) A sampling of MS scaling factors $F_{\textrm{MS}}$ [Eq.~(\ref{fms})] versus the defined width of the perimeter region ($x_0$), plotted for various boundary conditions, following the labeling convention of (a). The dashed black line marks $x_0=1$ $\upmu$m, the most common choice in our practice. This value strikes a balance between boundary condition insensitivity and computational ease.}
     \label{crosssection}
\end{figure}

The spatial distribution of electric field in the perimeter region can be written using separation of variables as $|E(x,y,z)|=C(y)f(x,z)$ in the limit of $x,z \rightarrow 0$.  This is because the electric field near a metal edge should have a local scaling property independent of distant electromagnetic boundary conditions.  Here $f(x,z)$ describes the edge scaling that can be applied to any cross section, independent of y. The actual form of $f(x,z)$ depends on material thicknesses and dielectric constants and is difficult to derive analytically.  However, we can compute $f(x,z)$ in a 2D cross-sectional electrostatic simulation of an electrode pad, which focuses on the metal edge and takes account of the actual thicknesses of each material [Fig.~\ref{areas}(b)].  The reduced dimensionality allows for accurate computation of the field inside the surface layer using sub-nm spatial resolution.  In this simulation we choose boundary conditions representative of the width of the pad ($w$) and the spatial separation between the opposing electrodes ($g$).  Although such a cross-sectional simulation does not accurately reflect the boundary condition in 3D space, as we already noted, $f(x,z)$ is independent of the distant boundary conditions as long as $x,z\ll g, w$.  As an illustration, $f(x,-t/2)$ is shown in Fig.~\ref{crosssection}(a) for a few very different values of $g$ and $w$.

For our devices with electrode pads typically 10 to 500 $\upmu$m in their smallest dimension, and separations on about the same scale, the above edge scaling function $f(x,z)$ is a very good approximation within the perimeter region for properly chosen $x_0$.  From $f(x,z)$ we can calculate the scaling factor $F_i$ based on the ratios of integrated field energy within the cross section:
\begin{equation}
F_{\textrm{MS}} = \frac{\int_{0}^{x_0}{dx}\int_{-t}^{0}{f^2(x,z)dz}}{\int_{x_0/2}^{x_0}{dx}\int_{-t}^{0}{f^2(x,z)dz}}
\label{fms}
\end{equation}
\begin{equation}
F_{\textrm{MA}} = \frac{\int_{0}^{x_0}{dx}\int_{h}^{h+t}{f^2(x,z)dz}+\int_{-t}^{0}{dx}\int_{0}^{h+t}{f^2(x,z)dz}}
{\int_{x_0/2}^{x_0}{dx}\int_{h}^{h+t}{f^2(x,z)dz}}
\end{equation}


Scaling factors $F_{\textrm{MS}}$ for various extents of the perimeter region are shown in Fig.~\ref{crosssection}(b).  We limit our method in the regime of $x_0\ll g,w$, where $F_{\textrm{MS}}$ is insensitive to the values of $g$ and $w$.  In practice, we use $x_0=1$ $\upmu$m for most of the pad structure (which are at least 10 $\upmu$m in width and separation).  
Inserting these simulated scaling factors into Eq.~(\ref{piper}) allows one to arrive at $p_{i,\textrm{per}}$.


\subsection{Surfaces associated with the junction leads}



\begin{figure}[t!]
  \centering
  \includegraphics[width=0.9\linewidth]{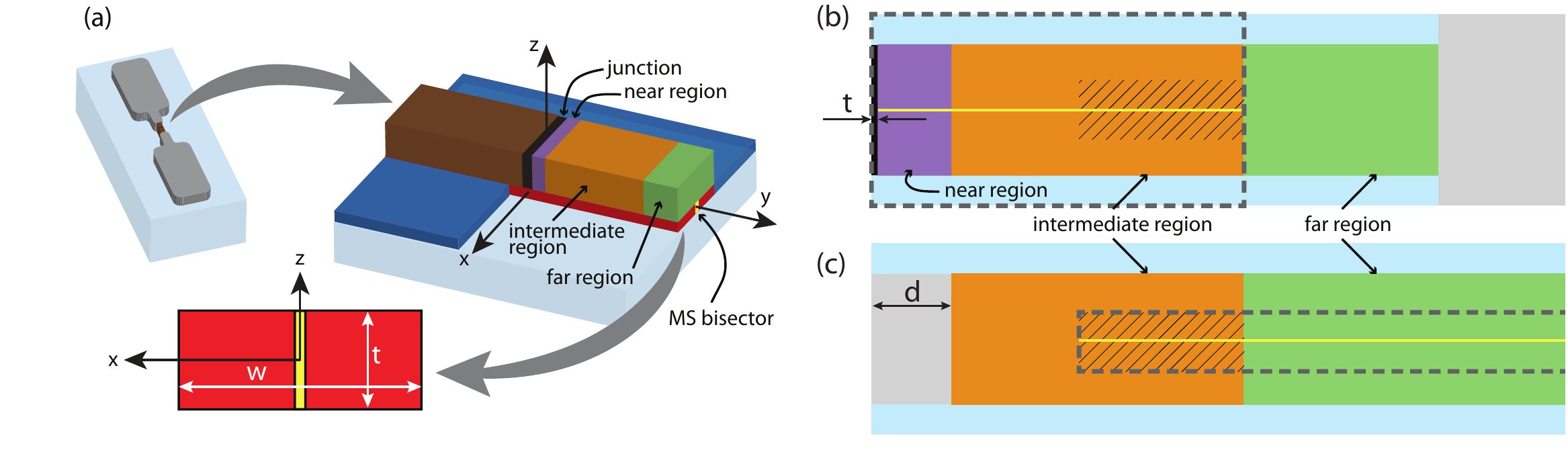}
  \caption{(a) A local, near-junction simulation is performed to evaluate lead participation ratios. Energy is integrated within cross-sections of the MS interface layer (red). Separately, energy measured along the bisector (yellow) is used for ``stitching'': comparing local and global simulations in convergent areas to establish a common energy scale. (b) In this top-view of the local simulation, the three regions of MS dielectric are color-coded in accordance with the conductor shading in (a). The junction (black) is explicitly simulated with thickness $t$. The near and intermediate regions, including adjacent SA dielectric, comprise the ``accurate'' area of the local simulation (dashed line bounding box); the far region is adversely affected by simulation boundaries and is ignored. The stitching region (diagonal lines) is located at the boundary of the intermediate and far regions, and at the edge of the local ``accurate'' area. The local simulation excludes some of the far region (gray). (c) In this zoomed-in top-view of the global simulation, the intermediate and far regions are defined, but the near area, extending distance $d$, is approximated as a lumped element. The accurate portion of the leads lies near their center (dashed line bounding box); energy near edges or the lumped-element area is divergent. The comparison of stitching extent (shaded area) along the bisector (yellow) to that of (b) determines the constant scaling between local and global electric fields. Within the stitching region, the cross-sectional energy density can be normalized by the bisector energy density to obtain an energy ratio $f(x,z)$ that is independent of $y$, useful for finding $p_{i,\textrm{far}}$ from Eq.~(\ref{pifar}).}
  \label{leadsim}
\end{figure}

A schematic of the Josephson junction and the leads is shown in Fig.~\ref{leadsim}(a), where $x$-axis and $y$-axis are defined perpendicular and parallel to the leads, respectively. We divide the surfaces associated with the junction leads into three regions based on distance from the junction: the near region ($|y|<1$ $\upmu$m), the intermediate region (1 $\upmu$m $<|y|<10$ $\upmu$m), and the far region ($|y|>10$ $\upmu$m). The surface participation ratios for these regions are denoted by $p_{i,\textrm{near}}$, $p_{i,\textrm{mid}}$ and $p_{i,\textrm{far}}$ respectively (as shown in Table~\ref{table}), where $i$ = MS, MA or SA.  The near region of the leads is not explicitly included in the global simulation.  The intermediate and far regions are included in the global simulation, but the surface integration of field energy does not converge due to the influence of edges.  A scaling factor solution akin to that in Section A demands $x,z\ll g,w$, but the lead is too narrow and too close to the junction to satisfy this.  We use a supplemental local 3D simulation of the junction leads as shown in Fig.~\ref{leadsim}(a), which includes the thicknesses of all materials, to compute the surface participation of all three regions surrounding the leads.

This high-resolution local simulation is performed by applying an electrostatic voltage potential between the pair of leads across the junction.  The boundary of the local simulation is set sufficiently far (typically 25 $\upmu$m) to ensure the calculated field distribution $\mathbf{E}_{\textrm{loc}}(x,y,z)$ in the the near and intermediate regions is not affected by the type of boundary condition used.  The overall magnitude of electric field in this local simulation is arbitrarily set by the imposed voltage, and must be rescaled by a constant $C$ to be consistent with the field scale of the global simulation from which $U_{\textrm{tot}}$ is obtained.

This constant $C$ can be determined by comparing $\mathbf{E}_{\textrm{loc}}(x,y,z)$ with the field distribution in the global simulation $\mathbf{E}_{\textrm{gbl}}(x,y,z)$ in a selected overlapping region (``stitching extent") where both simulations are reliable.  In particular, we choose the stitching extent as the center line of the leads in the 5 $\upmu\textrm{m}<|y|<10$ $\upmu$m region [Fig.~\ref{leadsim}(b)].  Such a choice avoids the numerical imprecision of the global simulation in areas close to the junction or the edges.  It also avoids any artificial boundary effects of the local simulation by remaining distant from the boundary.  We confirmed the two simulations show consistent spatial dependence over this stitching extent, $\mathbf{E}_{\textrm{gbl}}(0,y,0)\propto\mathbf{E}_{\textrm{loc}}(0,y,0)$, and the constant $C$ is computed from the ratio of the two. 


Surface participation ratios for the near and intermediate regions of the leads can then be immediately calculated by integrating $\mathbf{E}_{\textrm{loc}}(x,y,z)$ over the volume of interest.  For example,
\begin{equation}
p_{i,\textrm{near}} = C^2\iiint_{i,near}{\frac{\epsilon}{2}|\mathbf{E}_{\textrm{loc}}(x,y,z)|^2}dx\,dy\,dz / U_{\textrm{tot}}
\label{pinear}
\end{equation}
The surface participation ratios from the near and intermediate regions are expected to be independent of the design of the electrodes, and therefore show very little change among all the devices reported in this study (Table~\ref{table}).

On the other hand, Eq.~\ref{pinear} does not apply to lead energies in the far region, which is not fully included in the local simulation.  To calculate $p_{i,\textrm{far}}$ we adopt a separation-of-variables approach by noting that $|\mathbf{E}_{\textrm{glb}}(x,y,z)|$ = $|\mathbf{E}_{\textrm{glb}}(0,y,0)|f(x,z)$.  Here $f(x,z)$ describes the cross-sectional distribution of electric field in dimensionless units (normalized by the field magnitude at the center line of the lead) [Fig.~\ref{leadsim}(a)].  It can be obtained from the local simulation of the junction leads discussed above, which also confirms that $f(x,z)$ is independent of $y$ for $y\gg1$ $\upmu$m.  Therefore,
\begin{equation}
p_{i,\textrm{far}} = \int_{10\,\upmu\textrm{m}}^{y_{\textrm{far}}}\frac{\epsilon}{2}|\mathbf{E}_{\textrm{glb}}(0,y,0)|^2dy/U_{\textrm{tot}}\iint_{i}|f(x,z)|^2dx\,dz
\label{pifar}
\end{equation}
where the second integral effectively produces a constant factor that converts the electric field at a single point of the center line into energy per unit length along $y$.  This factor is equal to $7.5\times 10^{-15}$ m$^2$ for the typical lead width of 1 $\upmu$m.

\section{Fabrication methods}

Fabrication of qubits were performed using the Dolan bridge technique~\cite{Dolan1977} on 430 um thick c-plane EFG sapphire wafers. After cleaning in acetone and methanol, the wafer was spun with a bilayer of e-beam resist consisting of 550 nm of MMA EL13 and 70 nm of PMMA A3, then baked at 175 $^{\circ}$C.  A 13 nm aluminum film was then evaporated as an anti-charging layer for electron beam lithography.  Patterning of the qubit was done on a 100 kV VISTEC EBPG 5000+ e-beam writer.  The anti-charging layer was removed with TMAH, and the wafer was subsequently developed for 55 seconds in 1:3 MIBK:IPA followed by a 10 second rinse in IPA.  The wafer was then loaded into a Plassys e-beam evaporation system (MEB550S or UMS 300).  After a 40 W Ar/O$_2$ 3:1 plasma cleaning for 30 seconds, without breaking vacuum, a bi-layer of aluminum (20 nm and 60 nm) was deposited using double-angle evaporation. In between the two layers, the junction barrier was grown by thermal oxidation using a  Ar/O$_2$ 85\%/15\% mixture at 15 Torr for 12 minutes. Finally, the aluminum was capped with another oxide layer grown with the same mixture at 3 Torr for 10 minutes. After deposition, liftoff was performed in 60 $^{\circ}$C NMP for several hours, then rinsed with acetone and methanol. Prior to dicing, a layer of photoresist was spun on the wafer to protect the qubits. After dicing in an ADT ProVecturs 7100 dicer, the resist was removed by rinsing in acetone and methanol.

\begin{figure}[b!]
\includegraphics[width = 0.55\linewidth]{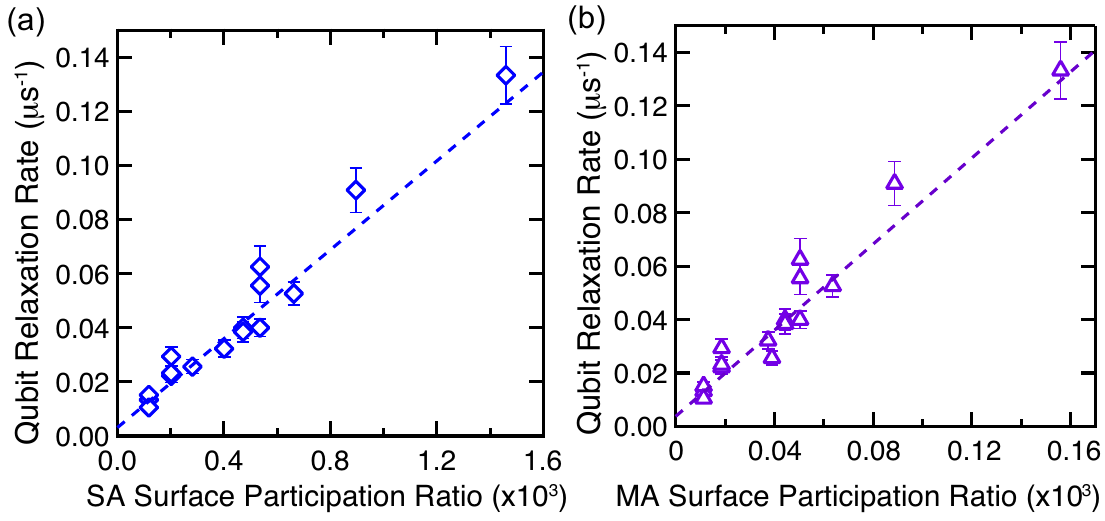}
\caption{Qubit relaxation rate ($1/T_1$) as a function of (a) substrate-air (SA), and (b) metal-air (MA) surface participation ratios, excluding contribution from the near region of the leads.  Dashed lines are linear fits to the data. 
}
\label{T1SAMA}
\end{figure}

\section{Qubit lifetime vs.~SA \& MA participation ratios}

In the main text, we presented the linear relationship between qubit relaxation rates ($1/T_1$) and the MS surface participation ratio ($p_{\textrm{MS}}$).  Since for all our qubit designs, MA, MS and SA surface participation ratios change approximately in proportion [Fig.~3(b) of the main text], qubit $1/T_1$ also shows similar linear relationship with $p_{\textrm{SA}}$ or $p_{\textrm{MA}}$, as shown in Fig.~\ref{T1SAMA}.  
Assuming SA or MA is the only lossy surface, linear fits to the two data sets indicate $\tan\delta_{SA}=2.2\times10^{-3}$ or $\tan\delta_{\textrm{MA}}=2.1\times10^{-2}$ respectively.   Since any of the three surfaces can be responsible for the qubit relaxation, these values, together with $\tan\delta_{\textrm{MS}}=2.6\times10^{-3}$ obtained from Fig.~3(a) of the main text, should be considered upper bounds for these dielectric surfaces.  Furthermore, since a combination of the three has to explain the strongly-correlated changes in $1/T_1$, we conclude $\tan\delta_{\textrm{MS}}+1.2\tan\delta_{\textrm{SA}}+0.1\tan\delta_{\textrm{MA}}=(2.6\pm0.1) \times10^{-3}$.  Linear fits in Fig.~\ref{T1SAMA}(a,b) also give residual (geometry-independent) relaxation rates of $3\pm1$ ms$^{-1}$ and $4\pm1$ ms$^{-1}$ respectively, consistent with the value obtained from Fig.~3(a).

\section{Estimate of surface participation in planar qubits}

In the main text, we placed a number of reported transmon qubits from literature on a diagram of $T_1$ versus $p_{\textrm{MS}}$ (Fig.~4) for comparison with devices in this study.  For most planar transmons included in the figure, we do not have complete knowledge of the geometric parameters related to all aspects of their design (\textit{e.g.} junction leads, coupling to the ground plane, device package, etc.) to perform a full-scale simulation.  However, since surface participation ratios for most planar transmons are dominated by capacitor pads with approximate translational symmetry (\textit{i.e.} having a longitudinal dimension much larger than the lateral dimension), we can estimate their $p_{\textrm{MS}}$ from cross-sectional simulations alone, similar to the previous work on CPW resonators~\cite{Sandberg2012, Wenner2011}.  Horizontal error bars of $\pm15\%$ represents uncertainties that can be caused by variations in parameters not captured in such a simulation.  The planar capacitors that we have simulated fall into three styles: interdigitated capacitor (IDC)~\cite{Chang2013, Houck2008, GeerlingsThesis, Riste2015}, coplanar waveguide (CPW)~\cite{Barends2013, Kelly2015} and coplanar capacitor (CPC)~\cite{Chow2014}.  All three styles can be simulated in settings similar to Fig.~\ref{areas}(b) with different choices of boundary conditions. 

In the context of translation-symmetric planar structures, the surface participation ratios are predominantly controlled by the width ($w$) of the capacitor electrodes and the gap ($g$) between them.  Assuming $w$ and $g$ are varied in proportion, the surface participation ratios are approximately inversely proportional to $w$ or $g$. (More rigorously, $p_i\propto \ln{(\frac{w}{t^{*}})}/w$, where $t^{*}$ is related to the thicknesses of the metal film ($h$) and the surface dielectric layers ($t$). For MS interface and for $h=80$ nm, $t=3$ nm, $t^{*}\approx8$ nm.) 
Therefore, it is convenient to express (inverse) surface participation ratios in the form of an effective length scale, $w_{eff}$.  We define $w_{eff}$, or ``effective IDC pitch width", of a qubit under study as the width (w) of an IDC structure (with g=w) with identical metal-substrate participation ratio.  This effective width has been used as the top axis in Fig.~4 of the main text, whose relationship to $p_{\textrm{MS}}$ is calibrated through cross-sectional simulations of IDC structures.
We also find the surface participation of CPW and CPC structures with $g=w$ are equivalent to IDC structures with $w_{\textrm{eff}}\approx 1.3g$ in both cases.  An advantage of using $w_{\textrm{eff}}$ is that surface participation ratios across different devices can be compared without assuming hypothetical thicknesses and dielectric constants of the surface dielectric layers.

The uncertainties of $T_1$ for these qubits reported from other institutions are based on the stated uncertainties, provided sample statistics, or the variations as a function of frequency (for frequency-tunable qubits).

\bigskip

\bibliography{SurfaceLoss_Supp_Final}

